\documentstyle[11pt,aaspp4]{article}

\received{15 December 1997}
\accepted{25 February 1998}

\slugcomment{Accepted for publication in Astrophysical Journal Letters}

\lefthead{O'Brien \& Cohen}
\righthead{On the jets in T Pyxidis}

\begin{document}

\title{On the existence of jets in the recurrent nova T Pyxidis
\altaffilmark{1}}

\author{T.J. O'Brien}
\affil{Astrophysics Research Institute, Liverpool John Moores University,
Liverpool L3 3AF, United Kingdom}

\and

\author{Judith G. Cohen}
\affil{Palomar Observatory, Mail Stop 105-24, California Institute of Technology, Pasadena, CA 91125.}

\altaffiltext{1}{Based on observations obtained at the W. M. 
Keck Observatory, which is operated jointly by the California Institute of
Technology and the University of California.}

\begin{abstract}
Recently, it has been claimed that the recurrent nova T Pyx exhibits
oppositely directed jets of ejecta apparent in features seen in
H$\alpha$ emission.  Here we demonstrate that these features are in
fact emission in the [N{\scriptsize II}] lines which lie either side
of H$\alpha$ and arise from the expanding shell associated with this
object rather than from collimated jets. We estimate an expansion
velocity along a line of sight through the centre of the shell of
about 500 km~s$^{-1}$.
\end{abstract}


\section{Introduction}

T Pyx is a recurrent nova with recorded outbursts in 1890, 1902, 1920,
1944 and 1966 (\cite{w87}). It is notable for possessing a bright
nebular shell extensively investigated by Shara et al.\ (1989) and
Shara et al.\ (1997).  Recently, Shahbaz et al.\ (1997 -- hereafter
S97) have presented optical spectroscopy of T Pyx in which they
identify emission components to the red and blue of H$\alpha$ (S$^+$
and S$^-$ respectively in their Figure 1) which they interpret as red-
and blue-shifted H$\alpha$ emission from oppositely directed jets.
These features occur at 6593 and 6539~$\rm{\AA}$ respectively implying
line-of-sight velocities of 1380 and $-1082$ km~s$^{-1}$.

\section{Observations and discussion}

On the night of November 21-22 1997 we obtained several spectra of T
Pyx with LRIS (the Low Resolution Imaging Spectrograph, Oke et al.\
1995) on the Keck II telescope on Hawaii. Two 400~second exposures
were made at slit position angles of 30$^\circ$ and 120$^\circ$.  The
600 grooves~mm$^{-1}$ grating was used with a slit width of 1.5~arcsec 
matching the seeing. The pixel scale was 0.2~arcsec 
and the spectral resolution 8.8~${\rm \AA}$.

Figure 1 shows a greyscale representation of the 2-dimensional
spectrum from position angle 30$^\circ$ clearly revealing the presence
of an expanding shell emitting in [N{\scriptsize II}] and
H$\alpha$ with signal to noise ranging from 10 to 70. 
The spectrum from position angle 120$^\circ$ is similar
although because the shell is not spherically symmetric (see Shara et
al.\ 1997) there are detailed differences.  In order to relate these
data to the spectra presented by S97, Figure 1 also includes the
spectrum obtained by summing along the slit. Although the spectral
resolution is lower than that of S97 it is still obvious that their
features S$^+$ and S$^-$ are in fact, respectively, the red-shifted
component of [N{\scriptsize II}]~6583 and the blue-shifted component
of [N{\scriptsize II}]~6548.

Further to this, by extracting the stellar continuum from our 2D
spectrum using Horne's optimal extraction method available in the
software package Figaro, and then subtracting this from the total
spectrum shown in Figure 1, we obtain a reasonable approximation to
the shell-only summed spectrum -- see Figure 2.  In this spectrum we
have indicated the rest wavelengths of H$\alpha$ and the two
[N{\scriptsize II}] lines together with the positions of the blue- and
red-shifted components arising from the front and back of a shell
expanding at a velocity of 530~km~s$^{-1}$. It is worth noting that
the component at about 6573~${\rm \AA}$ (arising from a combination of
blue-shifted [N{\scriptsize II}]~6583 and red-shifted H$\alpha$) is
clearly visible in the spectrum in Figure 1 of S97, whilst the
6555~${\rm \AA}$ component (from red-shifted [N{\scriptsize II}]~6548
and blue-shifted H$\alpha$) is blended into the blue wing of H$\alpha$
in their Figure 1.  It is difficult to estimate the expansion velocity 
from these spectra. Apart from the problems of contamination between
H$\alpha$ and [N{\scriptsize II}], the shell is clumpy and incomplete.
The velocity of 530~km~s$^{-1}$ is derived from the
 wavelength 6595~${\rm \AA}$ of the red-shifted [N{\scriptsize II}]~6583
line at the point where it crosses the stellar continuum. This
corresponds to the expansion velocity along a line of sight
through the centre of the shell and hence will be independent of slit
position angle. We estimate an uncertainty of $\pm 2$~${\rm \AA}$ on this
wavelength, equivalent to $\pm 90$~km~s$^{-1}$, 
as a result of the contamination by the stellar continuum and
of the spectral resolution. A better estimate would require more
kinematical data at higher resolution across the whole shell and a
plausible model for its structure. Note that the summed spectrum shown in 
Fig.\ 2 peaks shortward of 6595~${\rm \AA}$ because this particular feature is 
dominated by emission from a bright part of the shell 2-3~arcsec below the
star (see Fig.\ 1) which is at lower radial velocities. 

In conclusion we suggest there is little evidence to support the
existence of collimated jets in T Pyx. The data are however consistent
with the presence of a shell expanding at about 500~km~s$^{-1}$
and emitting more strongly in [N{\scriptsize II}] than in H$\alpha$ --
this is in broad agreement with the findings of Shara et al.\
(1989). However, more detailed modelling of the structure of the shell
and further higher spectral resolution observations are required in
order to reconcile this line-of-sight expansion velocity with the
upper limit on the velocity in the plane of the sky of 40~km~s$^{-1}$
derived from Hubble Space Telescope observations of the proper motion
of knots in the nebular shell (\cite{s97}).

\acknowledgments

We are grateful to M.\ Shara and C.\ Gill for useful discussions and
to an anonymous referee for helpful comments on the original version
of this paper.  The entire Keck/LRIS user community owes a huge debt
to Jerry Nelson, Gerry Smith, Bev Oke, and many other people who have
worked to make the Keck Telescope and LRIS a reality.  We are grateful
to the W. M. Keck Foundation, and particularly its late president,
Howard Keck, for the vision to fund the construction of the W. M. Keck
Observatory.

\clearpage

\begin{figure}
\vspace {20truecm}
\includegraphics{tpyxlog_all.eps}
\includegraphics{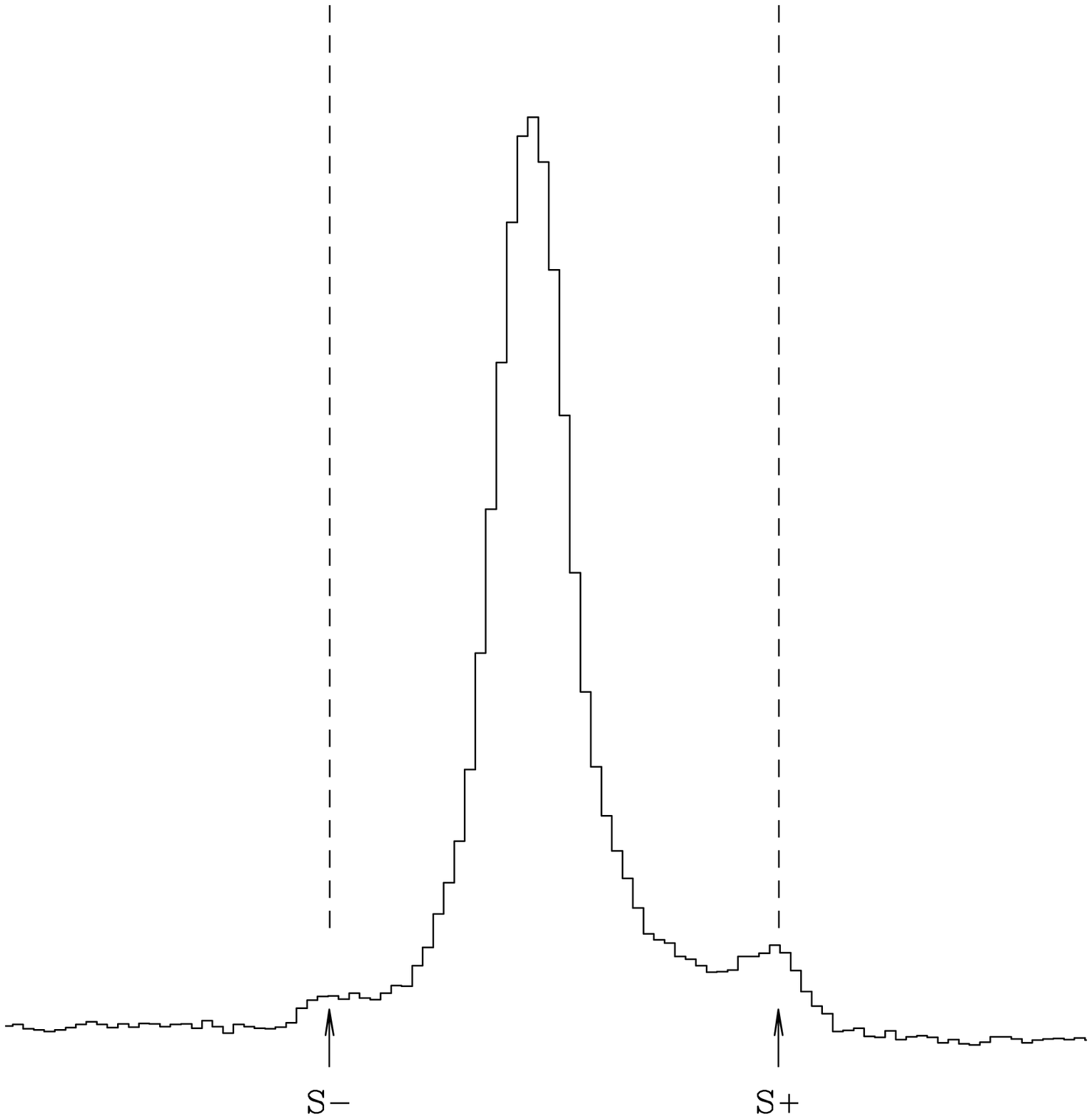}
\caption{The two-dimensional spectrum of T Pyx plotted as a logarithmic
grey-scale. Below it is the one-dimensional spectrum obtained by
summing in the spatial direction. The features referred to by Shahbaz
et al.\ (1997) as S$^+$ and S$^-$ are also indicated (at the
wavelengths taken from their paper) as is their origin in the
[N{\scriptsize II}] lines from the expanding shell.
\label{fig1}}
\end{figure}

\begin{figure}
\vspace {18truecm}
\includegraphics{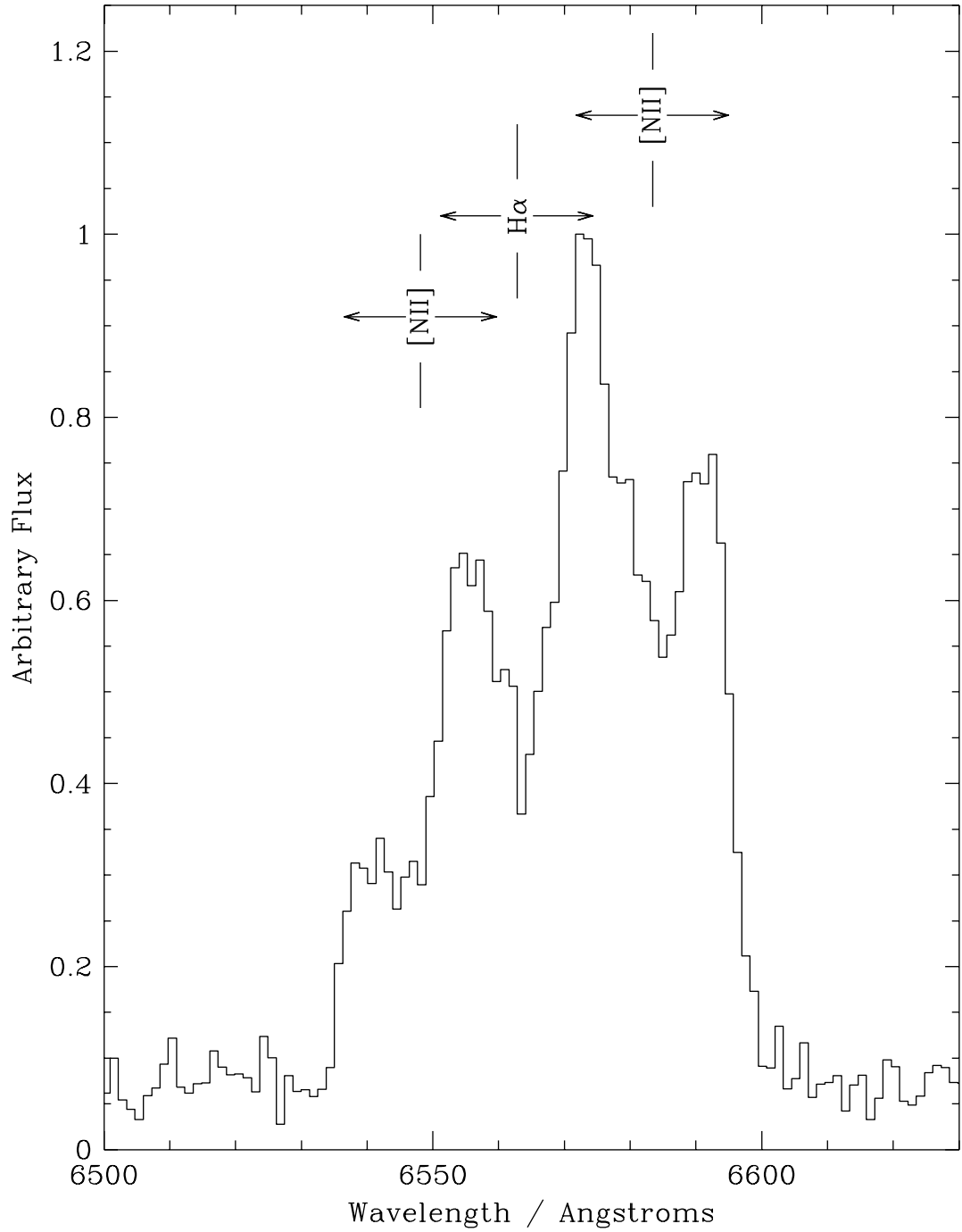}
\caption{The spectrum of the shell obtained by subtracting
the stellar contribution from the original data shown in Figure 1. The
rest wavelengths of H$\alpha$ and the two [N{\scriptsize II}] lines are 
indicated as are the wavelengths to which these lines would be 
red- and blue-shifted by an expansion along the line of sight of 
530~km~s$^{-1}$.
\label{fig2}}
\end{figure}

\end{document}